OPEN

# Growing interfaces uncover universal fluctuations behind scale invariance

Kazumasa A. Takeuchi[1], Masaki Sano[1], Tomohiro Sasamoto[2] & Herbert Spohn[3]

[1]Department of Physics, The University of Tokyo, 7-3-1 Hongo, Bunkyo-ku, Tokyo, 113-0033, Japan, [2]Department of Mathematics and Informatics, Chiba University, 1-33 Yayoi-cho, Inage, Chiba, 263-8522, Japan, [3]Zentrum Mathematik und Physik Department, TU München, D-85747 Garching, Germany.

Stochastic motion of a point – known as Brownian motion – has many successful applications in science, thanks to its scale invariance and consequent universal features such as Gaussian fluctuations. In contrast, the stochastic motion of a line, though it is also scale-invariant and arises in nature as various types of interface growth, is far less understood. The two major missing ingredients are: an experiment that allows a quantitative comparison with theory and an analytic solution of the Kardar-Parisi-Zhang (KPZ) equation, a prototypical equation for describing growing interfaces. Here we solve both problems, showing unprecedented universality beyond the scaling laws. We investigate growing interfaces of liquid-crystal turbulence and find not only universal scaling, but universal distributions of interface positions. They obey the largest-eigenvalue distributions of random matrices and depend on whether the interface is curved or flat, albeit universal in each case. Our exact solution of the KPZ equation provides theoretical explanations.

Scale invariance, *i.e.*, the absence of characteristic length and time scales, is a powerful concept in physics, which has provided simple and unified descriptions of natural phenomena. Prototypical examples are systems at equilibrium undergoing continuous phase transitions[1,2]. Ferromagnets and liquid-vapour systems, for example, become scale-invariant at the critical point because of the fractal like structure of spin configuration and of liquid/vapour patches, respectively. These phenomena are then ruled by a set of macroscopic laws without any specific scales, often manifested as power laws with universal characteristic exponents, despite the very different microscopic ingredients. With ample experimental evidence and deep theoretical understanding for systems at equilibrium[1,2], this universality has been a central subject of statistical mechanics since the mid 20th century.

Scale invariance is not restricted to thermal equilibrium. Great efforts have also been put into scale-invariant systems driven out of equilibrium, such as fully developed turbulence[3]. Here, we focus on growth phenomena[4–6] as one such nonequilibrium situation, which allows us to resolve very fine fluctuations both experimentally and theoretically as reported below, and thereby to address the fundamental issue of universality behind scale invariance. To start with an example, imagine a sheet of paper dipped to an ink suspension. One then observes the paper being wetted by the ink, typically with a rough interface that becomes even rougher as time elapses. Moreover, the interface looks scale-invariant as similar irregularities repeat at every scale. This can be quantified by introducing the interface height $h(x, t)$ and the width $w(l, t)$, which is simply the standard deviation of $h(x, t)$ over a length scale $l$ and thus measures the roughness of the interface. One then typically finds the following power law called the Family-Vicsek scaling[7]:

$$w(l,t) \sim t^\beta F\left(lt^{-1/z}\right) \sim \begin{cases} l^\alpha & \text{for } l \ll l_*, \\ l^\beta & \text{for } l \gg l_*, \end{cases} \quad (1)$$

with a scaling function $F$, two characteristic exponents $\alpha$ and $\beta$, the dynamic exponent $z \equiv \alpha/\beta$, and a crossover length scale $l_* \sim t^{1/z}$. Indeed, paper wetting[4] and many other phenomena such as fluid flow in porous media[8], bacterial colony growth[9], etc., as well as a large number of numerical models[4–6], have confirmed the scaling law (1). Moreover, numerical models have evidenced universality of the scaling exponents $\alpha$ and $\beta$, *e.g.*, $\alpha = 1/2$ and $\beta = 1/3$ for one-dimensional interfaces. This is theoretically well understood on the basis of a continuum equation proposed by Kardar, Parisi and Zhang (KPZ)[10,11] and constitutes the KPZ universality class[4–6].

Nevertheless, the universality has been quite elusive in experiments. To our knowledge, experiments on bacterial colony growth[9] and on paper combustion[12] are the only two that were able to show the KPZ exponents directly. Otherwise a few indirect indications of the KPZ scaling were reported in fracture surfaces[13,14], crystal facets[15] and gene segregation during bacterial growth[16], but an overwhelming majority of investigations[4–6] have reported values of $\alpha$ and $\beta$ inconsistent with the KPZ class. Major difficulties in the past experiments presumably lie in the existence of quenched disorder and long-ranged effective interactions, which are theoretically known to affect the universality[4–6,17]. Besides, statistical analysis has often been limited by rather moderate amount of available data under controlled conditions, a problem shared also by the above two experiments. Here, developing our previous work[18], we overcome all these difficulties and report a firm experimental observation of growing interfaces, which reveals universality of not only the scaling exponents but beyond, even in the distributions of shape fluctuations.

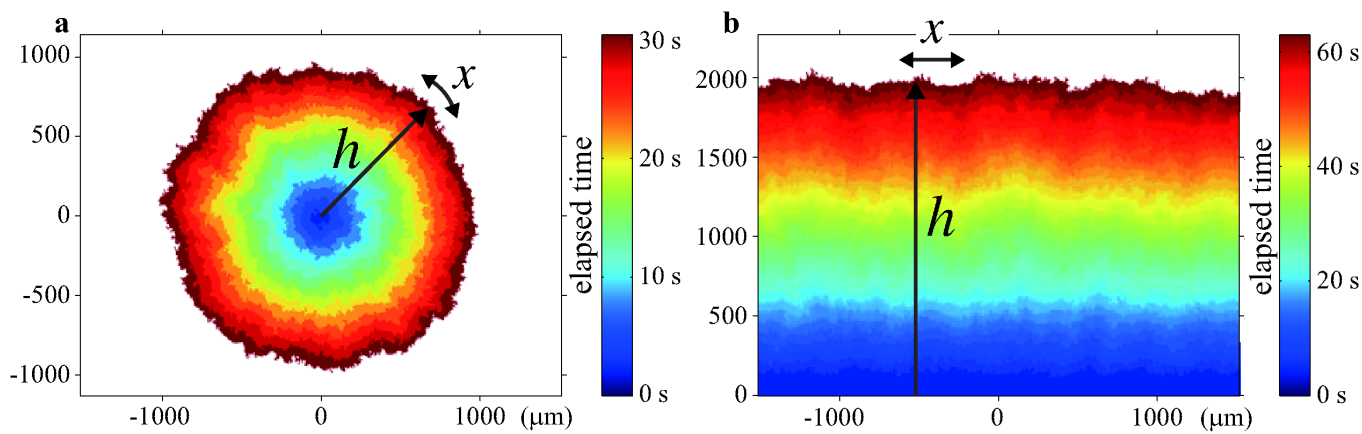

**Figure 1 | Growing DSM2 cluster with a circular (a) and flat (b) interface.** Binarised snapshots at successive times are shown with different colours. Indicated in the colour bar is the elapsed time after the laser emission. The local height $h(x, t)$ is defined in each case as a function of the lateral coordinate $x$ along the mean profile of the interface (a circle for a and a horizontal line for b). See also Supplementary Movies 1 and 2.

## Results

**Experimental evidence.** We study the convection of nematic liquid crystal, confined in a thin container and driven by an electric field[19,20], and focus on the interface between two turbulent states, called dynamic scattering modes 1 and 2 (DSM1 and DSM2)[20,21]. The latter consists of a large quantity of topological defects and can be created by nucleating a defect with a ultraviolet laser pulse. Whereas the generated DSM2 nucleus may disappear or lead to spatio-temporal intermittency at moderate applied voltages[21] around 22 V, for larger voltages it grows constantly, forming a compact cluster bordered by a moving rough interface (Fig.1a and Supplementary Movie 1).

This DSM2 growth has many advantages in the context of growing interfaces. It is a strictly local process and free from quenched disorder, because topological defects are simply elongated, split, and transported around by the turbulent flow, which has only short-ranged correlations in the DSM regimes and overwhelms cell heterogeneities. The experiment can be easily repeated under precisely controlled conditions. Moreover, we can realise flat interfaces as well (Fig.1b and Supplementary Movie 2), simply by shooting a line-shaped laser pulse through a cylindrical lens, so that we can study the geometry dependence of the interface fluctuations for the first time in experiments. In the following we report results obtained from 955 and 1128 realisations of circular and flat interfaces, respectively.

To characterise the roughness of the observed interface, we define the local height $h(x, t)$ along the moving direction of the interface, e.g. the local radius for the circular interface, as a function of the lateral coordinate $x$ (Fig.1). The interface width is then defined as $w(l,t) \equiv \left\langle \sqrt{\left\langle [h(x,t) - \langle h \rangle_l]^2 \right\rangle_l} \right\rangle$, where $\langle \cdots \rangle_l$ denotes the average over a segment of length $l$ and $\langle \cdots \rangle$ the average along the interface and all experimental realisations. Figure 2 displays the results, which confirm the Family-Vicsek scaling (1) in every aspect. The width $w(l, t)$ shows the power laws $w \sim l^\alpha$ for short lengths $l \ll l_* \sim t^{1/z}$ (Fig.2a,b) and $w \sim t^\beta$ for the largest $l$ (Fig.2c), measured from the overall width $W(t) \equiv \sqrt{\langle [h(x,t) - \langle h \rangle]^2 \rangle}$. In particular, these power laws evidence the exponents of the KPZ universality class for one-dimensional interfaces, namely $\alpha = 1/2$ and $\beta = 1/3$, regardless of the cluster shape. This is crosschecked in the insets of Fig.2ab, where we plot $w' \equiv w(l,t)t^{-\beta}$ against $l' \equiv lt^{-1/z}$ using the KPZ exponents and confirm that the data in the main panels collapse onto a single curve $w' = F(l')$ in agreement with equation (1). Therefore, we conclude that the DSM2 interface growth belongs to the KPZ class, demonstrating the robustness of the KPZ universality.

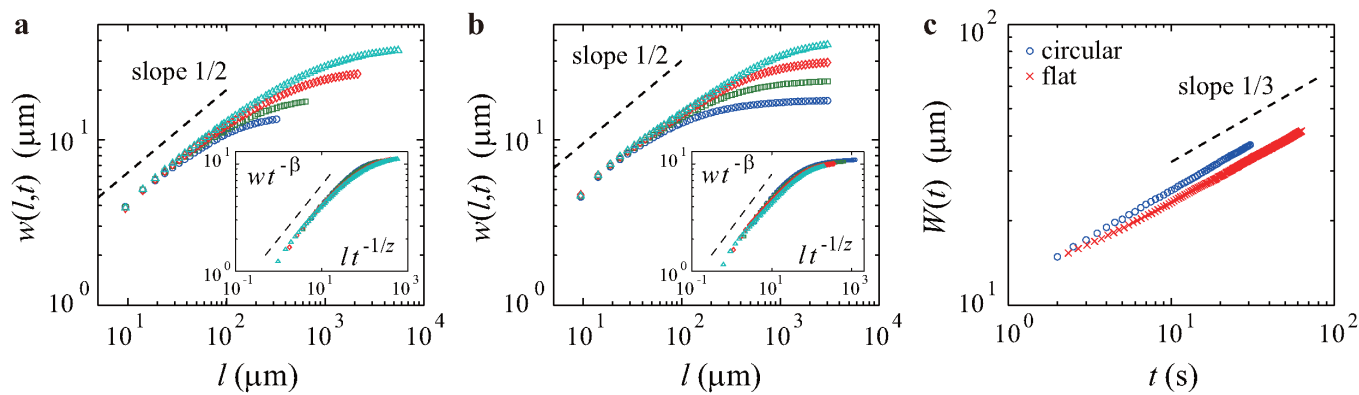

**Figure 2 | Family-Vicsek scaling.** a,b, Interface width $w(l, t)$ against the length scale $l$ at different times $t$ for the circular (a) and flat (b) interfaces. The four data correspond, from bottom to top, to $t = 2.0$ s, 4.0 s, 12.0 s and 30.0 s for the panel a and to $t = 4.0$ s, 10.0 s, 25.0 s and 60.0 s for the panel b. The insets show the same data with the rescaled axes. c, Growth of the overall width $W(t) \equiv \sqrt{\langle [h(x,t) - \langle h \rangle]^2 \rangle}$. The dashed lines are guides for the eyes showing the exponent values of the KPZ class.

Now we investigate the detailed form of the scale-invariant fluctuations. The Family-Vicsek scaling suggests that the height $h$ is composed of a deterministic linear growth term and a stochastic $t^{1/3}$ term:

$$h \simeq v_\infty t + (\Gamma t)^{1/3}\chi, \qquad (2)$$

with two parameters $v_\infty$, $\Gamma$ and the random amplitude $\chi$ which captures the fluctuations of the growing interace. We measure the values of the two parameters $v_\infty$ and $\Gamma$ from the experimental data (Supplementary Note 1) and make histograms of the rescaled height $\chi = (h - v_\infty t)/(\Gamma t)^{1/3}$ in Fig.3a. The result shows, surprisingly, clearly distinct distributions for the circular and flat interfaces (solid and open symbols), each of them not centred nor symmetric. To confirm we measure the second to fourth order cumulants of the height $h$, defined as $\langle h^2 \rangle_c \equiv \langle \delta h^2 \rangle$, $\langle h^3 \rangle_c \equiv \langle \delta h^3 \rangle$ and $\langle h^4 \rangle_c \equiv \langle \delta h^4 \rangle - 3\langle \delta h^2 \rangle^2$ with $\delta h \equiv h - \langle h \rangle$, and plot the skewness $\langle h^3 \rangle_c / \langle h^2 \rangle_c^{3/2}$ and the kurtosis $\langle h^4 \rangle_c / \langle h^2 \rangle_c^2$ in Fig.3b. Indeed, they have asymptotic values significantly different from zero unlike the Gaussian distribution, and distinct between the circular and flat interfaces.

In fact our experimental data in Fig.3a trace very precisely, without fitting, well-known distributions from a completely different context, namely the Tracy-Widom (TW) distributions of random matrix theory[22]. There are a few variants of the TW distributions. The data for the circular interfaces agree with the GUE TW distribution[23], which governs the largest eigenvalue distribution of complex hermitian random matrices (see below for details). In contrast those for the flat interfaces obey the real symmetric matrix counterpart, *i.e.* the GOE TW distribution[24]. The agreements are down to the scale $10^{-5}$, apart from a slight horizontal translation.

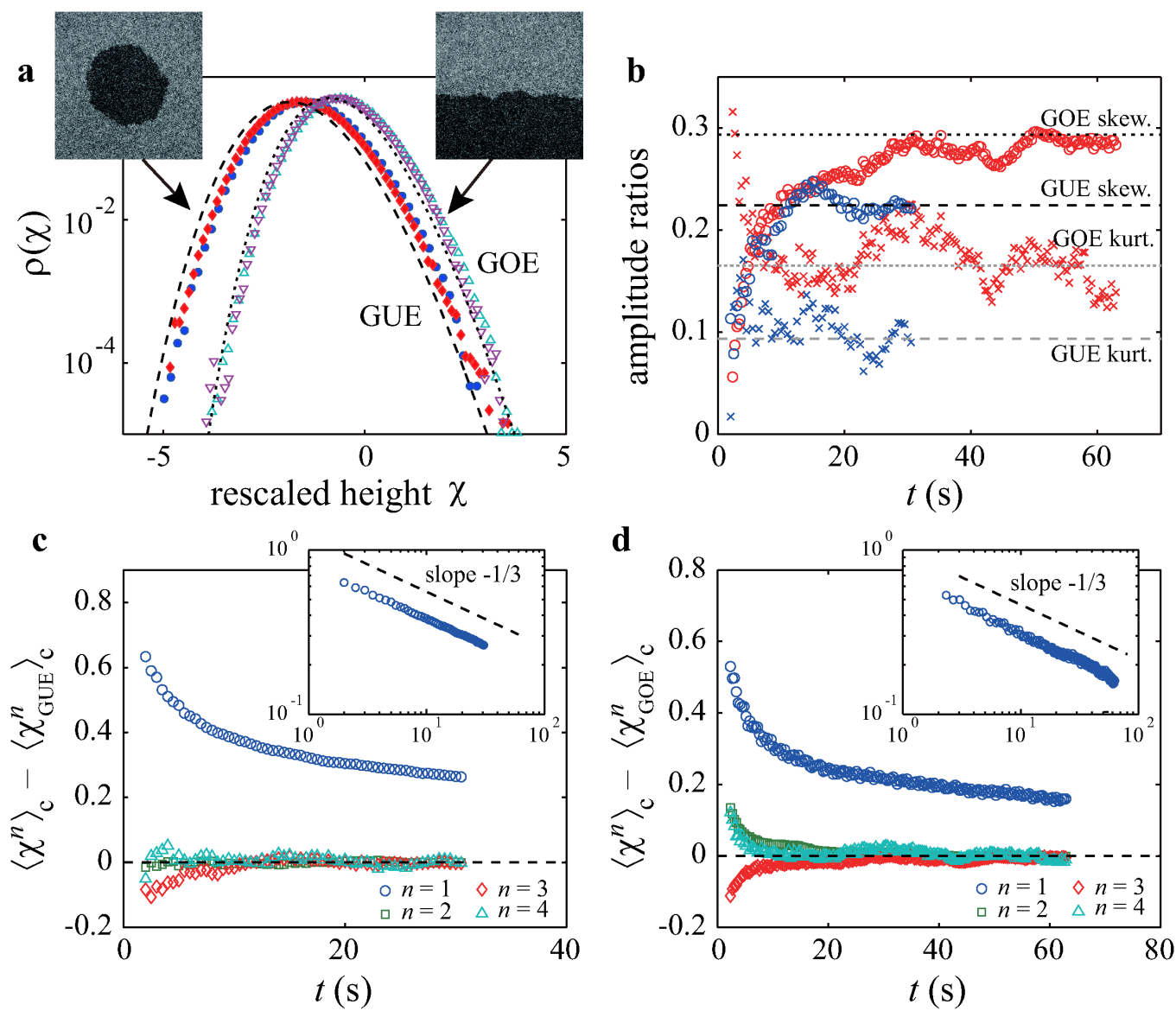

Figure 3 | Universal fluctuations. a, Histogram of the rescaled local height $\chi \equiv (h - v_\infty t)/(\Gamma t)^{1/3}$. The blue and red solid symbols show the histograms for the circular interfaces at $t = 10$ s and 30 s; the light blue and purple open symbols are for the flat interfaces at $t = 20$ s and 60 s, respectively. The dashed and dotted curves show the GUE and GOE TW distributions, respectively. Note that for the GOE TW distribution $\chi$ is multiplied by $2^{-2/3}$ in view of the theoretical prediction[31]. b, The skewness (circle) and the kurtosis (cross) of the distribution of the interface fluctuations for the circular (blue) and flat (red) interfaces. The dashed and dotted lines indicate the values of the skewness and the kurtosis of the GUE and GOE TW distributions[31]. c, d, Differences in the cumulants between the experimental data $\langle \chi^n \rangle_c$ and the corresponding TW distributions $\langle \chi^n_{\text{GUE}} \rangle_c$ for the ciruclar interfaces (c) and $\langle \chi^n_{\text{GOE}} \rangle_c$ for the flat interfaces (d). The insets show the same data for $n = 1$ in logarithmic scales. The dashed lines are guides for the eyes with the slope $-1/3$.

To elucidate this apparent deviation of the first moment, we plot in Fig.3cd the time series of the difference between the $n$th order cumulants of the measured $\chi$ and the theoretical values for the TW distributions. In both cases, the second to fourth order cumulants quickly converge to the corresponding TW values, whereas the first order cumulant $\langle\chi\rangle$, *i.e.* the mean, still has a significant deviation as suggested in the histograms. However, this deviation decreases in time, actually obeying a clear power law $\sim t^{-1/3}$ (the insets of Fig.3c,d), and vanishes in the $t \to \infty$ limit. To summarise, we have found that the interface fluctuations of the circular and flat interfaces precisely agree with the GUE and GOE TW distribution, respectively, at least up to the fourth order cumulant, with the pronounced finite-time effect proportional to $t^{-1/3}$ for the mean.

**Theoretical accounts.** Now we provide current theoretical understanding based on the KPZ equation (see also Supplementary Note 2 and a recent review[25,26]). First we illustrate how the TW distributions arise in random matrix theory. Let us consider an $N \times N$ complex hermitian matrix $A$ with matrix elements $A_{ij}$ satisfying $\bar{A}_{ij} = A_{ji}$. The matrix $A$ becomes random by drawing $A_{ij}$ from a Gaussian distribution. More precisely, we assume that the $A_{ij}$'s are independent with mean zero, $\langle A_{ij}\rangle = 0$, and variance $\langle |A_{ij}|^2 \rangle = N$. This statistical weight can be written more compactly as the Boltzmann factor $Z_N^{-1} \exp\left(-\frac{1}{2N}\text{Tr}A^2\right)$ with the normalizing constant $Z_N$, which defines the Gaussian unitary ensemble (GUE). $A$ has $N$ real eigenvalues. Since $A$ is random, so are the eigenvalues. Concerning the largest eigenvalue $\lambda_{\max}$, one finds

$$\lambda_{\max} \simeq 2N + N^{1/3}\chi_{\text{GUE}}, \qquad (3)$$

for large $N$. The random amplitude $\chi_{\text{GUE}}$ has the GUE TW distribution[23], which is shown by the left dashed line in Fig.3a. The cumulative distribution function of $\chi_{\text{GUE}}$ can be expressed as the determinant of an operator defined through the Airy kernel $K(x,y) = \int_0^\infty d\lambda \text{Ai}(x+\lambda)\text{Ai}(y+\lambda)$, where $x$ and $y$ range over the real line and Ai is the standard Airy function. Then

$$\text{Prob}(\chi_{\text{GUE}} \leq s) = \det(1 - P_s K P_s), \qquad (4)$$

where $P_s$ is the projection onto $[s, \infty)$. $P_s K P_s$ has the kernel $K(x, y)$ for $s \leq x, y$ and equals to 0 otherwise. The determinant in equation (4) is simply the product of the eigenvalues of the operator $1 - P_s K P_s$. For the Gaussian orthogonal ensemble (GOE), one follows the same construction, only now $A$ has to be real symmetric. The GOE TW distribution is again derived from its largest eigenvalue as in equation (3) with a determinant identity similar to equation (4)[24] and is indicated by the right dotted line in Fig.3a.

In their celebrated work[10], Kardar, Parisi, and Zhang proposed to model the interface growth through the stochastic evolution equation

$$\frac{\partial}{\partial t}h(x,t) = \frac{1}{2}\lambda\left(\frac{\partial}{\partial x}h(x,t)\right)^2 + \nu\frac{\partial^2}{\partial x^2}h(x,t) + \sqrt{D}\eta(x,t) \qquad (5)$$

with material dependent parameters $\lambda, \nu, D$ and white noise $\eta(x, t)$ representing the random nucleation events. In the flat case the initial height would be simply $h(x,0) = 0$, whereas for the droplet one would choose the parabola $h(x,0) = -(x/d)^2$ with small $d$ and focus only on the top part of the droplet, so that the interface may be described by a single-valued height $h(x, t)$. Studying behaviour of the equation under rescaling, one concludes[10,11] that the height fluctuations should grow as $t^{1/3}$ and the lateral correlation length as $t^{2/3}$, *i.e.*, $\alpha = 1/2$ and $\beta = 1/3$ as already mentioned above. On the other side, despite intense efforts, probability density distributions seemed to be unaccessible so far.

Exploiting lattice approximations[27] together with an intricate asymptotic analysis, we have obtained an exact solution of equation (5)[28,29]. It covers precisely the droplet growth of the experiment and provides a formula for the cumulative distribution function of $h(x, t)$. The distribution function, described through the determinant of a time-dependent operator, is found to have a structure comparable to equation (4) and to converge to the TW distribution in the limit of large $t$. This is consistent with our experimental result for the droplet growth (Fig.3a–c) and provides a strong theoretical evidence of the universality of scale-invariant fluctuations. From the exact solution one can further compute how fast the long time limit is approached. Indeed it turns out that the mean is the slowest mode and it approaches the TW mean as $t^{-1/3}$, which precisely agrees with our experimental observation in Fig.3c. For the flat case the exact solution of equation (5) is yet to be accomplished[34]. But for certain discrete models in the KPZ class, a mapping to a combinatorial problem has allowed a detailed analysis and the GOE TW distribution has been predicted[30–33], as indeed demonstrated by our experiment (Fig.3d).

## Discussion

It is remarkable that growing interfaces in a thin film have such profound statistics for the shape fluctuations. In addition to the unexpected link to the random matrix theory, it is rather counter-intuitive that, while in the KPZ class the scaling exponents are always the same, statistical properties of the height fluctuations do depend on the initial curvature of the interface. This is manifested in the two distinct but nevertheless universal height distributions, as clearly shown by our experiment. This result, together with the accompanying theory, provides strong evidence of the universality beyond the scaling laws, which underlies scale-invariant phenomena out of equilibrium.

## Methods

**Experimental setup and procedures.** The experimental setup is outlined in Supplementary Figure 1. A thin container of inner dimensions 16 mm × 16 mm × 12 $\mu$m is filled with liquid crystal $N$-(4-methoxybenzylidene)-4-butylaniline (MBBA) and 0.01 wt.% of tetra-$n$-butylammonium bromide. The liquid crystal molecules, initially aligned perpendicularly to the cell surfaces, strongly fluctuate when a sufficiently large ac voltage is applied between them and the first turbulent state DSM1 is reached. We then shoot 355 nm ultraviolet laser pulses to nucleate a topological defect in the liquid crystal orientation[21], which multiplies and forms the second turbulent state DSM2. For the experiments presented here, we shot two successive laser pulses of energy 6 nJ for the circular interfaces and 0.04 nJ/$\mu$m for the flat ones. The experiments were performed at temperature 25°C under the applied voltage 26 V at 250 Hz.

1. Stanley, H. E. *Introduction to Phase Transitions and Critical Phenomena* (Oxford Univ. Press, Oxford, 1987).
2. Henkel, M. *Conformal Invariance and Critical Phenomena* (Springer, Berlin, 1999).
3. Frisch, U. *Turbulence: The Legacy of A.N. Kolmogorov* (Cambridge Univ. Press, Cambridge, 1996).
4. Barabási, A.-L. & Stanley, H. E. *Fractal Concepts in Surface Growth* (Cambridge Univ. Press, Cambridge, 1995).
5. Meakin, P. The growth of rough surfaces and interfaces. *Phys. Rep.* **235**, 189–289 (1993).
6. Halpin-Healy, T. & Zhang, Y.-C. Kinetic roughening phenomena, stochastic growth, directed polymers and all that. Aspects of multidisciplinary statistical mechanics. *Phys. Rep.* **254**, 215–414 (1995).
7. Family, F. & Vicsek, T. Scaling of the active zone in the Eden process on percolation networks and the ballistic deposition model. *J. Phys. A* **18**, L75–L81 (1985).
8. Rubio, M. A., Edwards, C. A., Dougherty, A. & Gollub, J. P. Self-affine fractal interfaces from immiscible displacement in porous media. *Phys. Rev. Lett.* **63**, 1685–1688 (1989).
9. Wakita, J., Itoh, H., Matsuyama, T. & Matsushita, M. Self-affinity for the growing interface of bacterial colonies. *J. Phys. Soc. Jpn.* **66**, 67–72 (1997).
10. Kardar, M., Parisi, G. & Zhang, Y.-C. Dynamic scaling of growing interfaces. *Phys. Rev. Lett.* **56**, 889–892 (1986).
11. Forster, D., Nelson, D. R. & Stephen, M. J. Large-distance and long-time properties of a randomly stirred fluid. *Phys. Rev. A* **16**, 732–749 (1977).

12. Maunuksela, J. *et al.* Kinetic roughening in slow combustion of paper. *Phys. Rev. Lett.* **79**, 1515–1518 (1997).
13. Kertész, J., Horváth, V. K. & Weber, F. Self-affine rupture lines in paper sheets. *Fractals* **1**, 67–74 (1993).
14. Engøy, T., Måløy, K. J., Hansen, A. & Roux, S. Roughness of two-dimensional cracks in wood. *Phys. Rev. Lett.* **73**, 834–837 (1994).
15. Degawa, M., *et al.* Distinctive fluctuations in a confined geometry. *Phys. Rev. Lett.* **97**, 080601 (2006).
16. Hallatschek, O., Hersen, P., Ramanathan, S. & Nelson, D. R. Genetic drift at expanding frontiers promotes gene segregation. *Proc. Natl. Acad. Sci. USA* **104**, 19926–19930 (2007).
17. Csahók, Z., Honda, K. & Vicsek, T. Dynamics of suface roughening in disordered media. *J. Phys. A* **26**, L171–L178 (1993).
18. Takeuchi, K. A. & Sano, M. Universal fluctuations of growing interfaces: evidence in turbulent liquid crystals. *Phys. Rev. Lett.* **104**, 230601 (2010).
19. de Gennes, P. G. & Prost, J. The Physics of Liquid Crystals, *2nd ed* (Oxford Univ. Press, Oxford, 1993).
20. Kai, S. & Zimmermann, W. Pattern dynamics in the electrohydrodynamics of nematic liquid crystals. *Prog. Theor. Phys. Suppl.* **99**, 458–492 (1989).
21. Takeuchi, K. A., Kuroda, M., Chaté, H. & Sano, M. Experimental rezlization of directed percolation criticality in turbulent kiquid crystals. *Phys. Rev. E* **80**, 051116 (2009).
22. Mehta, M. L. *Random Matrices,* 3rd ed (Elsevier, Amsterdam, 2004).
23. Trach, C. & Widom, H. Level-spacing distributions and the Airy kenel. *Commun. Math. Phys.* **159**, 151–174 (1994).
24. Tracy, C. & Widom, H. On orthogonal and symplectic matrix ensembles. *Commun. Math. Phys.* **1777**, 727–754 (1996).
25. Kriecherbauer, T. & Krug, J. A pedestrian's view on interacting particle systems, KPZ universality and random matrices. *J. Phys. A* **43**, 403001 (2010).
26. Sasamoto, T. & Spohn, H. The 1 + 1-dimensional Kardar-Parisi-Zhang equation and its universality class. *J. Stat. Mech.* (2010) P11013.
27. Tracy, C. & Widom, H. Asymptotics in ASEP with step initial condition. *Commun. Math. Phys.* **290**, 129–154 (2009).
28. Sasamoto, T. & Spohn, H. One-dimensional Kardar-Parisi-Zhang equation: an exact solution and its universality. *Phys. Rev. Lett.* **104**, 230602 (2010).
29. Amir, G., Corwin, I. & Quastel, J. Probability distribution of the free energy of the continuum directed random polymer in 1+1 dimensions. *Commun. Pure Appl. Math.* **64**, 466–537 (2011).
30. Johansson, K. Shape fluctuations and random matrices. *Commun. Math. Phys.* **209**, 437–476 (2000).
31. Prähofer, M. & Spohn, H. Universal distributions for growth processes in 1+1 dimensions and random matrices. *Phys. Rev. Lett.* **84**, 4882–4885 (2000).
32. Baik, J. & Rains, E. M. in *Random Matrix Models and Their Applications* (eds Bleher, P. M. & Its, A. R.) vol. 40, 1–19 (Cambridge Univ. Press, Cambridge, 2001).
33. Borodin, A., Ferrari, P. L., Prähofer, M. & Sasamoto, T. Fluctuation properties of the TASEP with periodic initial configuration. *J. Stat. Phys.* **129**, 1055–1080 (2007).
34. After submission of our manuscript, an exact solution of the KPZ equation for the flat case was reported in Calabrese, P. & Le Doussal, P. An exact solution for the KPZ equation with flat initial conditions, arXiv:1104.1993v1 (2011).

### Acknowledgements

We are grateful to H. Chaté and M. Prähofer for fruitful discussions and to M. Prähofer for providing us with numerical values of the TW distributions. K.A.T. and M.S. acknowledge financial support by JSPS and MEXT (No.18068005). The work of T.S. is supported by KAKENHI (22740054).

### Author contributions

K.A.T. designed and performed the experiment and analyzed the data through discussions with M.S. The theoretical part is prepared mostly by T.S. and H.S. All the authors discussed the results.

### Additional information

**Supplementary Information** accompanies this paper at http://www.nature.com/scientificreports

**Competing financial interests:** The authors declare no competing financial interests.

**How to cite this article:** Takeuchi, K.A., Sano, M., Sasamoto, T. & Spohn, H. Growing interfaces uncover universal fluctuations behind scale invariance. *Sci. Rep.* **1**, 34; DOI:10.1038/srep00034 (2011).